\documentclass[conference]{IEEEtran}
\IEEEoverridecommandlockouts

\usepackage{cite}
\usepackage{amsmath,amssymb,amsfonts}
\usepackage{graphicx}
\usepackage{textcomp}
\usepackage{epstopdf}
\usepackage{amssymb}
\usepackage{xcolor}
\usepackage{amsthm}
\usepackage{fancyhdr}
\usepackage{verbatim}
\usepackage{flushend}
\usepackage[ruled]{algorithm2e} 
\usepackage{algpseudocode}
\usepackage[utf8]{inputenc}

\begin{document}
\title{Joint Transmit and Receive Beamforming  \\ 
	for Tri-directional Coil-Based Magnetic \\
	Induction Communications
\thanks{
    This work was supported in part by the Key Science and Technology Project of Ministry of Emergency Management of the People's Republic of China under grant 2024EMST131303, in part by the National Natural Science Foundation of China under grant 62401431, in part by Postdoctoral Fellowship Program of CPSF under grant GZC20241329, in part by the National Key Research and Development Program of China under Grant 2024YFB2908802, and in part by the Key R\&D Plan of Shaanxi Province under Grant 2024CY2-GJHX-29.\textit{(Corresponding author: Jianyu Wang.)}
}
}

\author{
	\IEEEauthorblockN{
		Jinyang Li\IEEEauthorrefmark{1}, 
		Jianyu Wang\IEEEauthorrefmark{1}, 
		Wenchi Cheng\IEEEauthorrefmark{1}, 
		Yudong Fang\IEEEauthorrefmark{2}
		and Wei Guo\IEEEauthorrefmark{2}} 
	\IEEEauthorblockA{\IEEEauthorrefmark{1}State Key Laboratory of Integrated Services Networks, Xidian University, Xi’an, China}
	\IEEEauthorblockA{\IEEEauthorrefmark{2}Ministry of Emergency Management Big Data Center, Beijing, China}
    E-mail: \textit{21009102228@stu.xidian.edu.cn, \{wangjianyu, wccheng\}@xidian.edu.cn,} \\
    \textit{fangyudong9713@ustc.edu, su27gw@163.com}
}

\maketitle

\begin{abstract}
In this paper, we enhance the omnidirectional coverage performance of tri-directional coil-based magnetic induction communication (TC-MIC) and reduce the pathloss with a joint transmit and receive magnetic beamforming method. An iterative optimization algorithm incorporating the transmit current vector and receive weight matrix is developed to minimize the pathloss under constant transmit power constraints. We formulate the mathematical models for the mutual inductance of tri-directional coils, receive power, and pathloss. The optimization problem is decomposed into Rayleigh quotient extremum optimization for transmit currents and Cauchy-Schwarz inequality-constrained optimization for receive weights, with an alternating iterative algorithm to approach the global optimum. Numerical results demonstrate that the proposed algorithm converges within an average of 13.6 iterations, achieving up to 54\% pathloss reduction compared with equal power allocation schemes. The joint optimization approach exhibits superior angular robustness, maintaining pathloss fluctuation smaller than 2 dB, and reducing fluctuation of pathloss by approximately 45\% compared with single-parameter optimization methods.
\end{abstract}


\section{Introduction}

Magnetic induction communication (MIC) is an emerging and promising research field due to the advantage of low pathloss, stable channel response and low propagation delay \cite{kisseleff2018survey}. Typical applications include soil condition monitoring, earthquake prediction, and communication in mines and tunnels \cite{akyildiz2002wireless,liu2024frequency}. However, traditional magnetic induction communication based on single-coil systems suffers from uneven magnetic field distribution and limited communication angles. Tri-directional coils are proposed to eliminate the sensitivity of antenna angles \cite{guo2015channel}. A complex underwater magnetic induction communication channel model was proposed in \cite{guo2017multiple}, using tri-directional coils to eliminate the sensitivity of MI antenna angle changes.

It has been shown that employing multiple transmit or receive coils can significantly enhance the performance of magnetic induction communications by constructively combine the magnetic fields \cite{xu2016investigation,lee2017precise}. The technique is termed magnetic beamforming \cite{yang2016magnetic}. Under the impetus of Multiple-Input Multiple-Output (MIMO) technology \cite{zhao2017artificial}, magnetic beamforming can optimize the power allocation for transmit coils and the receive weights for receive coils, thereby concentrating maximizing the receive power and improving efficiency. The magnetic induction communication system with multiple transmit coils and/or multiple receive coils has been studied in \cite{yoon2011investigation,ahn2012effect,jadidian2014magnetic,moghadam2015multiuser}. The magnetic induction communication system with two transmit coils and one receive coil is studied in \cite{yoon2011investigation,ahn2012effect}. Recently, an “Magnetic MIMO” charging system \cite{jadidian2014magnetic} can charge a phone inside a user's pocket 40 cm away from the array of transmit coils, independently of the phone's orientation. For a MIC system with multiple receive coils, the load resistances of the receive coils are jointly optimized in \cite{moghadam2015multiuser} to minimize the total transmit power and address the “near-far” fairness problem. Deploying multiple transmit coils can help focus the magnetic fields on the receive coils \cite{jadidian2014magnetic}, in a manner analogous to beamforming in far-field wireless communications \cite{gershman2010convex}. \cite{yang2016magnetic} designs the magnetic beamforming from a signal processing and optimization perspective.

In\cite{wang2022multi}, a Multi-frequency Resonating Compensation (MuReC) coil-based simultaneous wireless power transfer and magnetic induction communication system is proposed. Also, a beamforming scheme is proposed in \cite{wang2022multi} to minimize transmit power and ensure network performance. The authors of \cite{wang2023backscatter} proposed a backscatter based bidirectional full-duplex magnetic induction communication (BFMC), and formulate the joint magnetic beamforming and time allocation optimization problem to minimize the average energy consumption.

As shown in Fig.~\ref{fig1}, we consider a scenario where both the transmitters and receivers are equipped with the tri-directional coils. The currents of the transmit coils and the signal weights of the receive coils can be adjusted to achieve magnetic beamforming. We formulate an optimization problem, which involves designing the transmit currents and receive weights to minimize system pathloss under the constraint of constant transmit power. We use an iterative optimization algorithm to approach the optimal solution. Simulation results show that, compared with uniform distribution, magnetic beamforming significantly reduces the pathloss and enhance the omnidirectional coverage
performance of tri-directional coil-based magnetic induction communication (TC-MIC).

\section{System Model}
\begin{figure*}[t!]
	\centering
	\includegraphics[scale=0.57]{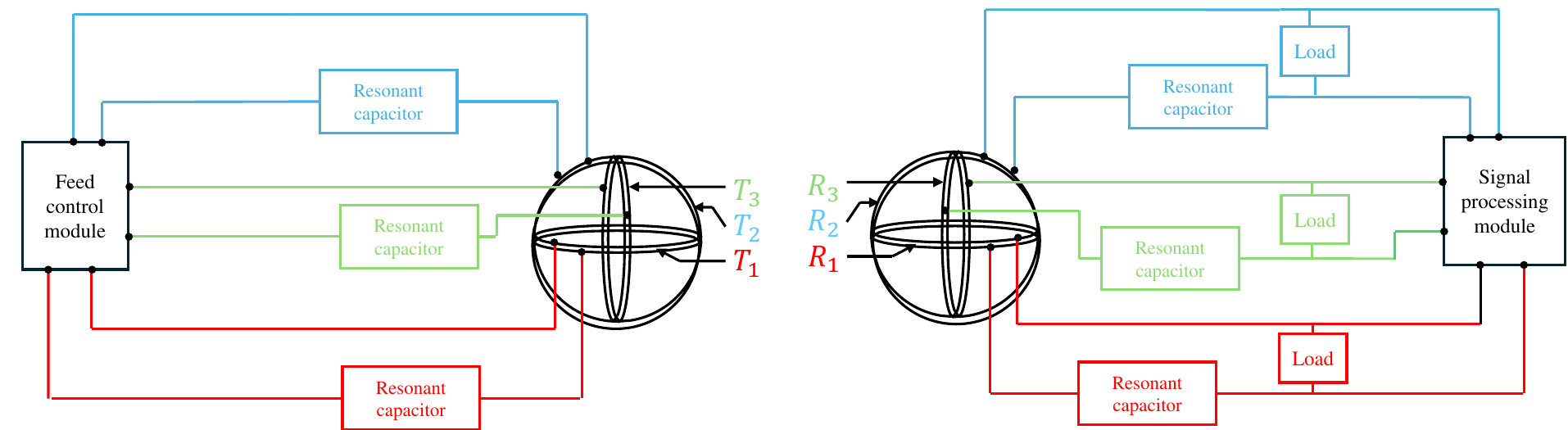}
	\caption{Equivalent circuit of the TC-MIC.}
	\label{fig1}
\end{figure*}

The equivalent circuit of the TC-MIC system is shown in Fig. 1, where both the transmitters and receivers are equipped with the tri-directional coils. The  transmit currents are distributed through a feed control module. To optimize the signal quality at the receive coils, the voltage from the tri-directional coils is combined in a weighted ratio via a signal processing module. The currents in the transmit coils $T_1$, $T_2$, and $T_3$ are respectively expressed as
	\begin{align}
		I_1 &= I_{m1} \cdot \sin(\omega t), \tag{1a} \\
		I_2 &= I_{m2} \cdot \sin(\omega t), \tag{1b} \\
		I_3 &= I_{m3} \cdot \sin(\omega t), \tag{1c}
	\end{align}
where $I_{m1}$, $I_{m2}$, and $I_{m3}$ represent the current amplitudes of the three transmit coils. To prevent mutual interference between the internal coils of the tri-directional coils, each coil is equipped with a capacitor such that the resonant frequency of the capacitor-coil combination is maintained at  $\omega$. The circuit connection diagram of the tri-directional transceiver coils is illustrated in the Fig.~\ref{fig1}.

We define the mutual inductance matrix $\mathbf{M}$ as a $3\times3$ matrix, where the element $M_{ij}$ represents the mutual inductance between the transmit coil $T_i$ and the receive coil $R_j$:
\begin{equation}
	\mathbf{M}^T = 
	\begin{bmatrix}
		M_{11} & M_{21} & M_{31} \\
		M_{12} & M_{22} & M_{32} \\
		M_{13} & M_{23} & M_{33}
	\end{bmatrix}.
	\tag{2}
\end{equation}
The mutual inductance between transmit coil $T_1$ and receive coil $R_n$ (where $n=1,2,3$) can be expressed as:
\begin{equation}
	\begin{split}
		M_{1n} = \frac{\mu_0 N_t N_r S_t S_r}{4\pi r^5} \bigg( & 3xz \cos \alpha_n + 3yz \cos \beta_n \\
		& + (2z^2 - x^2 - y^2) \cos \gamma_n \bigg).
	\end{split}
	\tag{3}
\end{equation}
For the transmit coil \( T_2 \), according to the coordinate transformation: \( Z \rightarrow X,\ \ X \rightarrow Y, \ \ Y \rightarrow Z \), the normal vectors of the receive coils become: \(\mathbf{n}_1 = (\cos\gamma_1, \cos\alpha_1, \cos\beta_1)\), \(\mathbf{n}_2 = (\cos\gamma_2, \cos\alpha_2, \cos\beta_2)\), \(\mathbf{n}_3 = (\cos\gamma_3, \cos\alpha_3, \cos\beta_3)\).

Therefore, the mutual inductance between the transmit coil $T_1$ and the receive coil $R_n$ can be expressed as:
\begin{equation}
	\begin{split}
		M_{2n} =\frac{\mu_0 N_t N_r S_t S_r}{4\pi r^5} \bigg( & \, 3xy \cos\beta_n + 3xz \cos\alpha_n \\
		& + (2x^2 - y^2 - z^2) \cos\gamma_n \bigg).
	\end{split}
	\tag{4}
\end{equation}
Similarly, for the transmit coil $T_3$, we have: $Y\rightarrow X,\ \ Z\rightarrow Y,\ \ X\rightarrow Z$, so its mutual inductance with the receive coil $R_n$ can be expressed as:
\begin{equation}
	\begin{split}
		M_{3n} =\frac{\mu_0 N_t N_r S_t S_r}{4\pi r^5} \bigg( & \, 3yz \cos\gamma_n + 3xy \cos\alpha_n \\
		& + (2y^2 - x^2 - z^2) \cos\beta_n \bigg).
	\end{split}
	\tag{5}
\end{equation}

The receive voltage vector $\mathbf{E}=[E_1,E_2,E_3]^T$ can be expressed as:
\begin{equation}
	\mathbf{E} = j\omega \mathbf{MI},
	\tag{6}
\end{equation}
where $\mathbf{I}=[I_1,I_2,I_3]^T$ represents the transmit current vector, and  $\omega$ is the signal angular frequency. Let the receive weight matrix be a diagonal matrix $\mathbf{S}=\mathrm{diag}(s_1,s_2,s_3)$, then the receive power $P_r$ is:
\begin{equation}
	P_r = \frac{Z_L \omega^2}{(Z_r + Z_L)^2} \cdot \mathbf{I}^H \mathbf{M}^H \mathbf{S}^H \mathbf{S} \mathbf{M} \mathbf{I}
	\tag{7}.
\end{equation}

The transmit power is the sum of the power of each transmit coil, expressed as
\begin{equation}
	P_t = \mathbf{I}^H R_t \mathbf{I},
	\tag{8}
\end{equation}
where $R_t$ is the resistance of the transmit coil. Therefore, the pathloss is
\begin{equation}
	L = -10 \log_{10} \left( \frac{\mathbf{I}^H \mathbf{M}^H \mathbf{S}^H \mathbf{S} \mathbf{M} \mathbf{I}}{\mathbf{I}^H R_t \mathbf{I}} \cdot \frac{Z_L \omega^2}{(Z_r + Z_L)^2} \right).
	\tag{9}
\end{equation}

\section{Problem Formulation}
We formulate the optimization problem of power distribution at the transmit coils and the weight matrix $\mathbf{S}$ at the receive coils: 
\begin{align}
	\underset{I,S}{\text{maximize}} \quad & \frac{\mathbf{I}^T \mathbf{M}^T \mathbf{S}^T \mathbf{S} \mathbf{M} \mathbf{I}}{\mathbf{I}^T \mathbf{I}} \tag{10a} \\
	\text{s.t.} \quad & \|\mathbf{S}\|^2 = 1; \tag{10b} \\
	& \mathbf{I}^T \mathbf{I} = \frac{P_0}{R_t}. \tag{10c}
\end{align}
Under the actual constraints of the system, we design the current vector $\mathbf{I}$ flowing through the transmit coils and the weight matrix $\mathbf{S}$ at the receive coils to minimize the pathloss of the system. We consider the following two constraints: power constraint, to ensure the quality of the transmit signal, the total power at the receive coils should be equal to the preset power $P_0$; and the receive weights normalization constraint, that is, the sum of the squares of the receive weights should be 1.

To solve this optimization problem, we first fix the transmit current $\mathbf{I}$ and optimize the weight matrix $\mathbf{S}$; then fix $\mathbf{S}$ and optimize the transmit current $\mathbf{I}$; and finally alternate the optimization of parameters to approach the optimal solution.

\section{Optimal Solution}
We first fix the receive weights diagonal matrix $\mathbf{S}$. Since the result of $\mathbf{M}^T\mathbf{S}^T\mathbf{SM}$ is definitely a Hermitian matrix, the objective function is the Rayleigh quotient, and the optimization problem can be transformed into:
\begin{align}
	\underset{I}{\text{maximize}} \quad & \frac{\mathbf{I}^T \mathbf{Q} \mathbf{I}}{\mathbf{I}^T \mathbf{I}}, \quad \mathbf{Q} = \mathbf{M}^T \mathbf{S}^T \mathbf{S} \mathbf{M} \tag{11a} \\
	\text{s.t.} \quad & \mathbf{I}^T \mathbf{I} = \frac{P_0}{\mathbf{R}_t}. \tag{11b}
\end{align}

According to the properties of the Rayleigh quotient, the maximum value of the objective function is determined by the maximum eigenvalue of matrix $\mathbf{Q}$, and the corresponding eigenvector is the optimal current.

Let the eigenvalues of the matrix $\mathbf{Q}$ be $\lambda_i$, with the largest eigenvalue being $\lambda_{max}$, and the corresponding unit eigenvector be $\boldsymbol{v}^*$. From the power constraint condition: $\mathbf{I}^T\mathbf{I}=P_0/R_t$, the optimal transmit current is:
\begin{equation}
	\mathbf{I}^* = \sqrt{\frac{P_0}{\mathbf{R}_t}} \boldsymbol{v}^*\tag{12}.
\end{equation}
Next, fix the transmit current $\mathbf{I}$ and optimize the receive weights $\mathbf{S}$. At this time, since the denominator of the objective function is a constant, the original problem is simplified to:
\begin{align}
	\underset{S}{\text{maximize}} \quad & \mathbf{I}^T \mathbf{M}^T \mathbf{S}^T \mathbf{S} \mathbf{M} \mathbf{I} \tag{13a} \\
	\text{s.t.} \quad & \|\mathbf{S}\|^2 = 1. \tag{13b}
\end{align}

Let $\mathbf{m}_i$ be the i-th row of matrix $\mathbf{M}$, define vector $\mathbf{a}=\left(|\mathbf{m}_\mathbf{1}\mathbf{I}|,|\mathbf{m}_\mathbf{2}\mathbf{I}|,|\mathbf{m}_\mathbf{3}\mathbf{I}|\right)$, vector $\mathbf{s}=(s_1,s_2,s_3)$, then the objective function can be expressed as:
\begin{equation}
	J(S) = \sum_{i=1}^{3} s_i^2 (\mathbf{m}_i \mathbf{I})^2 = \|\mathbf{a} \mathbf{s}^T\|_2^2\tag{14}.
\end{equation}
According to the Cauchy-Schwarz inequality, for any vectors $\mathbf{u},\mathbf{v}\in\mathbb{R}^n$, we have:
\begin{equation}
	\|\mathbf{u} \mathbf{v}^T\|_2^2 \leq \|\mathbf{u}\|_2^2 \|\mathbf{v}\|_2^2 \tag{15}.
\end{equation}

(15) holds if and only if $\mathbf{u}$ and $\mathbf{v}$ are linearly dependent.
Let $\mathbf{u}=\mathbf{s}, \mathbf{v}=\mathbf{a}$, we get:
\begin{align}
	\begin{split}
	\sum_{i=1}^{3} s_i^2 (\mathbf{M}_i \mathbf{I})^2 &\leq \left( \sum_{i=1}^{3} s_i^2 \right) \left( \sum_{i=1}^{3} (\mathbf{m}_i \mathbf{I})^2 \right) \\
	&= \sum_{i=1}^{3} (\mathbf{m}_i \mathbf{I})^2,
	\end{split}
	\tag{16}
\end{align}
if and only if the weights $s_i$ are proportional to the receive signal magnitude $|\mathbf{M}_\mathbf{i}\mathbf{I}|$, the objective function takes the maximum value. Therefore, the optimal receive weights are:
\begin{equation}
	\mathbf{S}^* = \text{diag} \left(
	\begin{array}{c}
		\frac{\|\mathbf{m}_1\|}{\sqrt{\sum_{j=1}^{3} (\mathbf{m}_j \mathbf{I})^2}}, \\
		\frac{\|\mathbf{m}_2\|}{\sqrt{\sum_{j=1}^{3} (\mathbf{m}_j \mathbf{I})^2}}, \\
		\frac{\|\mathbf{m}_3\|}{\sqrt{\sum_{j=1}^{3} (\mathbf{m}_j \mathbf{I})^2}}
	\end{array}
	\right) .\tag{17}
\end{equation}

After obtaining the above two individually optimized closed-form solutions, Algorithm 1 can be designed to iteratively optimize and approximate the optimal solution.

\begin{algorithm}
	\caption{Iterative Optimization Algorithm}
	\begin{algorithmic}[1] 
		\Require physical parameters of the transmit and receive coils, position information of the receive coils, and system operating frequency
		\Ensure Optimal solution $\mathbf{S}^*$ and $\mathbf{I}^*$
		
		\State Set $n=0$, choose $\mathbf{S}=\mathbf{S}_{(\mathbf{0})}$, calculate the mutual inductance matrix $\mathbf{M}$ between the coils based on the formula.
		\State Fix the receive weights, calculate the maximum eigenvalue of matrix $\mathbf{Q}=\mathbf{M}^T\mathbf{S}^T\mathbf{SM}$ and the corresponding unit eigenvector $\boldsymbol{v}^*$.
		\State Calculate the optimal transmit current 
		\begin{equation*}
			\mathbf{I}_\mathbf{(n)} = \sqrt{\frac{P_0}{R_t}} \boldsymbol{v}^*.
		\end{equation*}
		\State Fix the transmit current, update the optimal receive weights  
		\begin{align*}
	\mathbf{S}_\mathbf{(n)} &= \mathbf{S}^\mathbf{*} \\
	&= \text{diag} \left( \frac{|\mathbf{m}_\mathbf{1} \mathbf{I}|}{\sqrt{\sum_{j=1}^{3} (\mathbf{m}_\mathbf{j} \mathbf{I})^2}}, \right. \\
	&\qquad \left. \frac{|\mathbf{m}_\mathbf{2} \mathbf{I}|}{\sqrt{\sum_{j=1}^{3} (\mathbf{m}_\mathbf{j} \mathbf{I})^2}}, \frac{|\mathbf{m}_\mathbf{3} \mathbf{I}|}{\sqrt{\sum_{j=1}^{3} (\mathbf{m}_\mathbf{j} \mathbf{I})^2}} \right).
\end{align*}
		\State Calculate the pathloss 
		\begin{equation*}
			\begin{split}
				L^{(n)} &= \frac{Z_L \omega^2}{(Z_r + Z_L)^2} \frac{\mathbf{I}_{(n)}^T \mathbf{M}^T \mathbf{S}_{(n)}^T \mathbf{S}_{(n)} \mathbf{M} \mathbf{I}_{(n)}}{\mathbf{I}_{(n)}^T R_t \mathbf{I}_{(n)}}.
			\end{split}
		\end{equation*}
		\State $n = n + 1$
		\State Repeat steps 2-6
		\State If $L^{(n)}-L^{\left(n-1\right)}\le\delta$, exit the loop
		\State Output $\mathbf{S}^\ast=\mathbf{S}_{(n)}$, $\mathbf{I}^\ast=\mathbf{I}_{(n)}$
		
	\end{algorithmic}
\end{algorithm}

\section{Simulation Result}
 We set the number of turns of the transmit and receive coils $N_r=N_t=10$; the radius of the coils: 0.1m; the amplitude of the transmit current $I_m=2V$; the resistance of the coil material per unit length $R_0=0.01\Omega/m$; the center point P of the receive coils is at coordinates P(1,1,1.5). We assume the normal vector of the receive coil $R_1$ in the plane is $\mathbf{n}_1=(\sqrt{1-\cos^2\alpha},0,\cos\alpha)$. When $\mathbf{n}_1\neq[1,0,0]$, a right-handed basis is generated through vector operation $\mathbf{c}=[1,0,0]-\mathbf{n}_1$, and another coil normal vector is obtained using the cross product of vectors.

\begin{equation}
	\mathbf{n}_2 = \frac{\mathbf{c}}{\|\mathbf{c}\|}, \mathbf{n}_3 = \mathbf{n}_1 \times \mathbf{n}_2 \tag{18},
\end{equation}
where $\mathbf{n}_1=[1,0,0]$, $\mathbf{n}_2=[0,1,0], n3=[0,0,1]$.
\subsection{Selection of Convergence threshold}
\begin{figure}[htbp]
	\centering
	\includegraphics[scale=0.43]{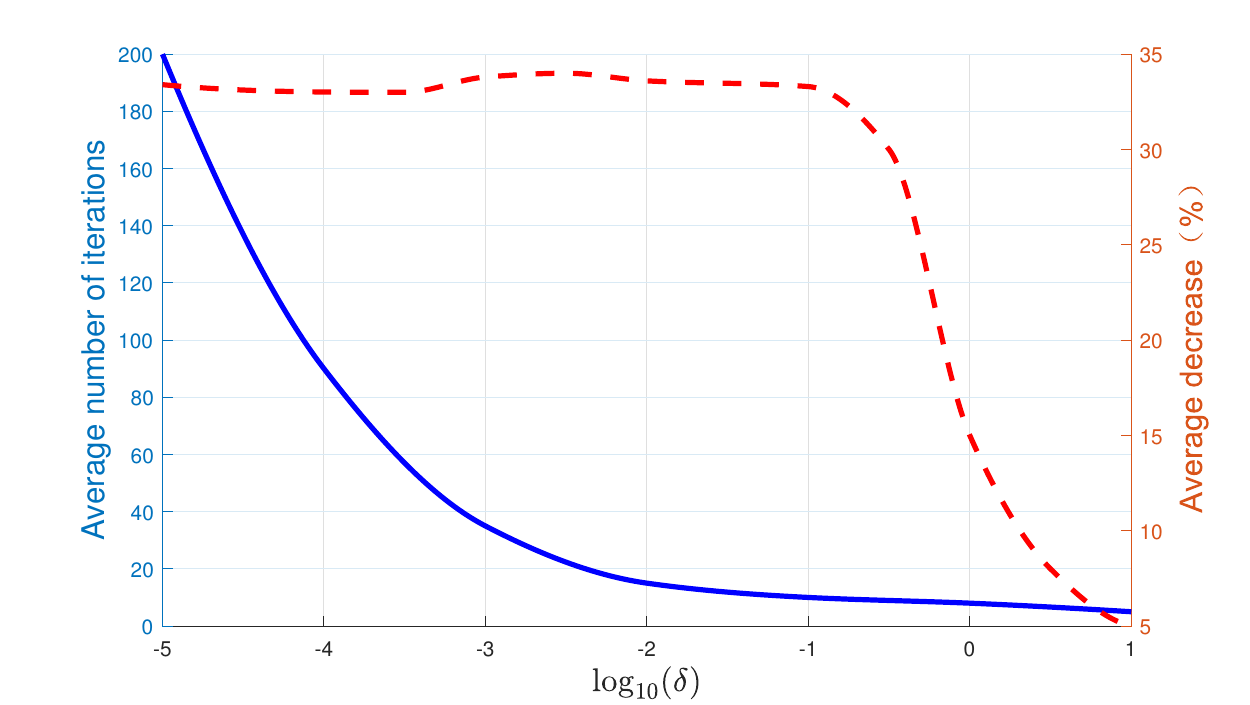}
	\caption{The relationships between the convergence threshold $\delta$ and the average optimization amplitude, as well as the average number of optimizations.}
	\label{fig2}
\end{figure}

From Fig.~\ref{fig2}, it can be seen that if the convergence threshold is too small, the number of iterations in the system  increases sharply, and may even fail to converge. When the convergence threshold $\delta$ increases, the number of iterations decreases, approaching zero. When $\delta$ is less than $3\times{10}^{-1}$, the average reduction of the system fluctuates around 34\%. When $\delta$ is greater than $3\times{10}^{-4}$, the average reduction of the system decreases, indicating that the number of iterations is insufficient at this point, leading to a loss of system performance. Therefore, while ensuring that the number of iterations is sufficient, it is also necessary to keep the number of iterations in the system as small as possible. To this end, we choose the convergence threshold $\delta=2.5\times{10}^{-2}$. According to Fig.~\ref{fig2}, the average reduction of the system is 34.7\%, and the average number of iterations is 13.6. In the subsequent simulations, this convergence threshold will be used.

\subsection{Optimization Process and Performance Comparison}
Through iterative optimization, the system can dynamically adjust the current matrix at the transmit coils and the weight matrix at the receive coils to adapt to communication requirements at different angles, thereby minimizing pathloss. The algorithm has different optimization effects at different angles: when $\alpha=1$, the pathloss decreases from 18.8 dB to 8.6 dB, with a reduction of about 54\%; when $\alpha=\pi$, the pathloss decreases from 11.4 dB to 8.6 dB, with a reduction of about 24\%. The algorithm converges at the 7th and 9th iterations, respectively when $\alpha=1$ and $\alpha=\pi$. At other angles, the iterative process usually converges in about 10 iterations, indicating a relatively small number of iterations. This shows that when the orientation of the receive coils is adjusted, the system can complete algorithm optimization in time to maintain normal communication between the transmit and receive coils.

\begin{figure}[htbp]
	\centering
	\includegraphics[scale=0.6]{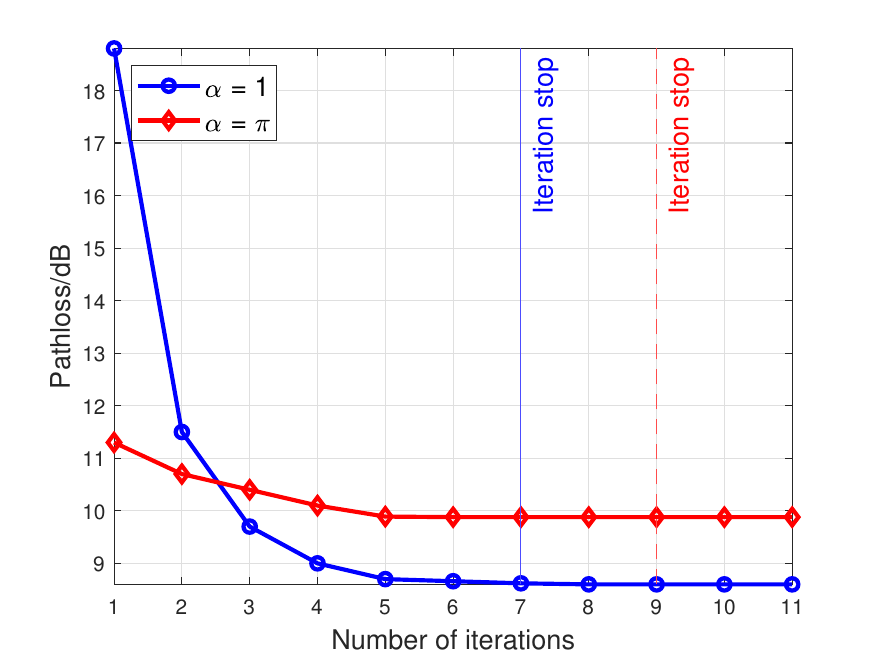}
	\caption{Iterative optimization process.}
	\label{figure3}
\end{figure}

\begin{figure}[htbp]
	\centering
	\includegraphics[scale=0.6]{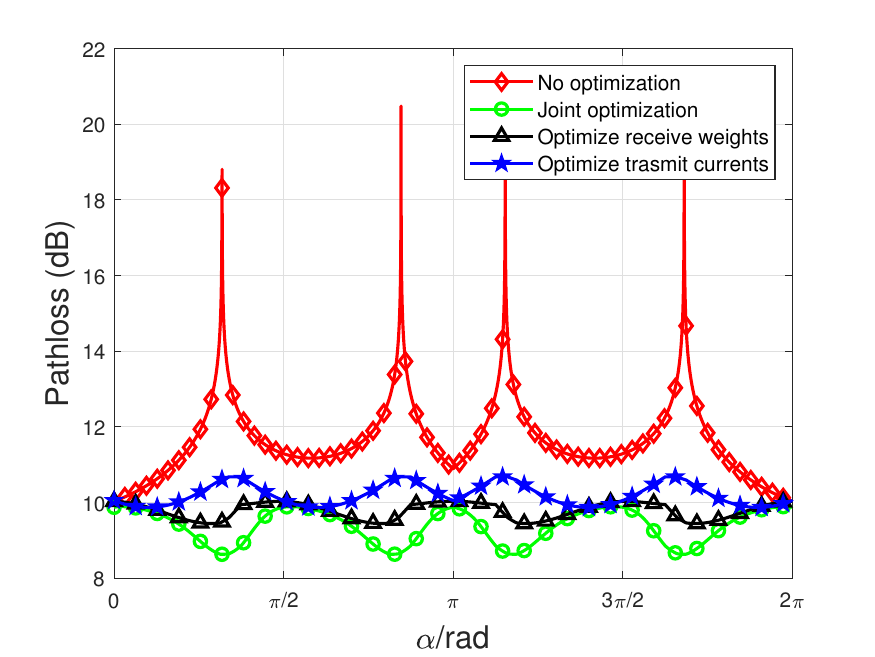}
	\caption{Pathloss fluctuation with $\alpha$ under different optimization methods.}
	\label{fig4}
\end{figure}

To verify the weak dependence of the optimized communication system on the orientation of the coils, we fix the communication distance and simulate the system where the coil rotates. We compare the pathloss of the optimized system with that of the tri-directional coil-based communication system with equal power allocation and obtain Fig.~\ref{fig4}.

It is shown in Fig.~\ref{fig4} that after introducing the proposed optimization algorithm, the system performance has been significantly improved: the overall pathloss of the iterative optimization method is stable within the 8.5-10 dB, and the system pathloss does not fluctuate drastically with changes in angle; if only the transmit currents are optimized, although the pathloss has a similar numerical range to the iterative optimization, there is greater loss near $\pi/4$ and $3\pi/4$; if only the receive weights are optimized, the pathloss is stable throughout the 9.5-10 dB range, with a maximum fluctuation of about 0.5 dB, showing strong robustness to changes in spatial angles. It is worth noting that the loss difference at $\alpha=\pi/2$, $\pi$, $3\pi/2$, and $\pi$ for the three optimization methods is less than 0.3 dB, indicating that different optimization objectives may achieve similar optimization effects under certain spatial constraints.

\section{Conclusion}
In this paper we proposed a joint transmit and receive beamforming scheme for tri-directional coil-based magnetic induction communication. With an iterative optimization algorithm for the transmit current vector and receive weight matrix, the system pathloss was successfully reduced by 34.7\%. Numerical results demonstrate that: (1) the proposed alternating optimization algorithm converges within an average of 13.6 iterations, significantly improving system adaptability to spatial orientations; (2) the optimized system maintains pathloss within 8.5-10 dB, achieving up to 54\% performance improvement compared to conventional equal power allocation schemes; (3) the joint optimization approach exhibits superior angle robustness with fluctuation of pathloss reduced by approximately 45\% compared to single-parameter optimization methods.

\bibliographystyle{IEEEtran}
\bibliography{reference_ucom.bib}

\begin{thebibliography}{10}
\providecommand{\url}[1]{#1}
\csname url@samestyle\endcsname
\providecommand{\newblock}{\relax}
\providecommand{\bibinfo}[2]{#2}
\providecommand{\BIBentrySTDinterwordspacing}{\spaceskip=0pt\relax}
\providecommand{\BIBentryALTinterwordstretchfactor}{4}
\providecommand{\BIBentryALTinterwordspacing}{\spaceskip=\fontdimen2\font plus
\BIBentryALTinterwordstretchfactor\fontdimen3\font minus
  \fontdimen4\font\relax}
\providecommand{\BIBforeignlanguage}[2]{{%
\expandafter\ifx\csname l@#1\endcsname\relax
\typeout{** WARNING: IEEEtran.bst: No hyphenation pattern has been}%
\typeout{** loaded for the language `#1'. Using the pattern for}%
\typeout{** the default language instead.}%
\else
\language=\csname l@#1\endcsname
\fi
#2}}
\providecommand{\BIBdecl}{\relax}
\BIBdecl

\bibitem{kisseleff2018survey}
S.~Kisseleff, I.~F. Akyildiz, and W.~H. Gerstacker, ``Survey on advances in
  magnetic induction-based wireless underground sensor networks,'' \emph{IEEE
  Internet of Things Journal}, vol.~5, no.~6, pp. 4843--4856, 2018.

\bibitem{akyildiz2002wireless}
I.~F. Akyildiz, W.~Su, Y.~Sankarasubramaniam, and E.~Cayirci, ``Wireless sensor
  networks: a survey,'' \emph{Computer networks}, vol.~38, no.~4, pp. 393--422,
  2002.

\bibitem{liu2024frequency}
G.~Liu, ``Frequency-switchable routing protocol for dynamic magnetic
  induction-based wireless underground sensor networks,'' \emph{IEEE Journal of
  Selected Areas in Sensors}, vol.~1, pp. 1--8, 2024.

\bibitem{guo2015channel}
H.~Guo, Z.~Sun, and P.~Wang, ``Channel modeling of mi underwater communication
  using tri-directional coil antenna,'' in \emph{2015 IEEE Global
  Communications Conference (GLOBECOM)}.\hskip 1em plus 0.5em minus 0.4em\relax
  IEEE, 2015, pp. 1--6.

\bibitem{guo2017multiple}
------, ``Multiple frequency band channel modeling and analysis for magnetic
  induction communication in practical underwater environments,'' \emph{IEEE
  Transactions on Vehicular Technology}, vol.~66, no.~8, pp. 6619--6632, 2017.

\bibitem{xu2016investigation}
B.~Xu and Y.~Li, ``Investigation of surface wave propagation along a multi-coil
  wireless power transfer system,'' \emph{Microwave and Optical Technology
  Letters}, vol.~58, no.~9, pp. 2261--2265, 2016.

\bibitem{lee2017precise}
S.~B. Lee, C.~Lee, and I.~G. Jang, ``Precise determination of the optimal coil
  for wireless power transfer systems through postprocessing in the smooth
  boundary representation,'' \emph{IEEE Transactions on Magnetics}, vol.~53,
  no.~6, pp. 1--4, 2017.

\bibitem{zhao2017artificial}
N.~Zhao, Y.~Cao, F.~R. Yu, Y.~Chen, M.~Jin, and V.~C. Leung, ``Artificial noise
  assisted secure interference networks with wireless power transfer,''
  \emph{IEEE Transactions on Vehicular Technology}, vol.~67, no.~2, pp.
  1087--1098, 2017.

\bibitem{yoon2011investigation}
I.-J. Yoon and H.~Ling, ``Investigation of near-field wireless power transfer
  under multiple transmitters,'' \emph{IEEE Antennas and Wireless Propagation
  Letters}, vol.~10, pp. 662--665, 2011.

\bibitem{ahn2012effect}
D.~Ahn and S.~Hong, ``Effect of coupling between multiple transmitters or
  multiple receivers on wireless power transfer,'' \emph{IEEE Transactions on
  Industrial Electronics}, vol.~60, no.~7, pp. 2602--2613, 2012.

\bibitem{jadidian2014magnetic}
J.~Jadidian and D.~Katabi, ``Magnetic mimo: How to charge your phone in your
  pocket,'' in \emph{Proceedings of the 20th annual international conference on
  Mobile computing and networking}, 2014, pp. 495--506.

\bibitem{moghadam2015multiuser}
M.~R.~V. Moghadam and R.~Zhang, ``Multiuser charging control in wireless power
  transfer via magnetic resonant coupling,'' in \emph{2015 IEEE International
  Conference on Acoustics, Speech and Signal Processing (ICASSP)}.\hskip 1em
  plus 0.5em minus 0.4em\relax IEEE, 2015, pp. 3182--3186.

\bibitem{gershman2010convex}
A.~B. Gershman, N.~D. Sidiropoulos, S.~Shahbazpanahi, M.~Bengtsson, and
  B.~Ottersten, ``Convex optimization-based beamforming,'' \emph{IEEE Signal
  Processing Magazine}, vol.~27, no.~3, pp. 62--75, 2010.

\bibitem{yang2016magnetic}
G.~Yang, M.~R.~V. Moghadam, and R.~Zhang, ``Magnetic beamforming for wireless
  power transfer,'' in \emph{2016 IEEE International Conference on Acoustics,
  Speech and Signal Processing (ICASSP)}.\hskip 1em plus 0.5em minus
  0.4em\relax IEEE, 2016, pp. 3936--3940.

\bibitem{wang2022multi}
J.~Wang, W.~Cheng, W.~Zhang, W.~Zhang, and H.~Zhang, ``Multi-frequency access
  for magnetic induction-based swipt,'' \emph{IEEE Journal on Selected Areas in
  Communications}, vol.~40, no.~5, pp. 1679--1691, 2022.

\bibitem{wang2023backscatter}
J.~Wang, W.~Cheng, W.~Zhang, and H.~Zhang, ``Backscatter based bidirectional
  full-duplex magnetic induction communications,'' \emph{IEEE Transactions on
  Communications}, vol.~71, no.~11, pp. 6258--6271, 2023.

\end{thebibliography}

\end{document}